# Application of Response Surface Method and Genetic Algorithm in the Design of High-Efficiency Prototype Vehicle

Paras Singh[1], Harshit Gupta[1*], Ojas Vinayak[1], Aryan Tyagi[1]

[1] Department of Mechanical Engineering, Delhi Technological University, New Delhi, India

*hgh1602@gmail.com

## ABSTRACT

Breakthroughs in aerodynamic optimization have made it possible to develop efficient modes of transport with lesser exploitation of valuable resources. This makes it crucial for technical professionals such as engineers and scientists to understand the methodologies behind carrying out such optimizations. A common approach towards improving the aerodynamic properties of a vehicle is to alter its physical shape, which has concurrently been a very strenuous process given the time consumed to remodel the vehicle for each simulation process. This research aims to tackle this problem by using intelligent techniques to automate the step-by-step process of remodeling the car and arriving at a final optimized solution with a significantly lower drag coefficient, a quantity used to measure the amount of drag force acting on a vehicle. This is achieved by assigning particular parameters to ensure guided improvement of the airfoil in a process known as parametrization, followed by implementing a response surface methodology primarily to circumvent the strenuous task of performing a large number of CFD simulations by employing surrogate models to generate a response surface between selected independent variables. Further, evolutionary algorithms such as Genetic Algorithm have gained momentum in the optimization studies carried out during product design by selecting the optimum parameters from the available design spaces on the basis of natural evolution. The proposed method of optimization has been successfully implemented on a prototype vehicle with an improvement of 26.6% and 51.1% in the drag coefficient and drag area respectively.

# 1 Introduction

Understanding airflow dynamics in the automotive sector is crucial for effective aerodynamic design within technological constraints [1]. A substantial amount of fuel can be saved provided a vehicle is aerodynamically compliant. Vehicle optimization during the design process aims to enhance desired aerodynamic properties, with minimizing drag through improved body design being a prominent strategy. This improves fuel efficiency, increases top speeds, and passenger comfort. Crucial developments in the past have led to the technological advancements in aerodynamic design optimization that we employ today.

Past studies have demonstrated the efficiency of optimization frameworks in various fields [2,3,4]. For example, Kaveh et al. [5] emphasize the benefits of employing the CS algorithm in the creation of two-dimensional steel frames. Their research demonstrates that Levy flights outperformed Gaussian or uniform distributions, which are frequently employed by alternative metaheuristic algorithms when it comes to navigating extensive design spaces.

Furthermore, work has been done on researching optimization algorithms and their roles in the aerodynamic optimization of various components, specifically, the use of sensitivity algorithms has also been studied in relation with computational fluid dynamics to optimize shapes [6], and the use of benchmarking optimization algorithms for wing design optimization [7,8].

The present study adds to the existing literature by exploring the application of genetic algorithms, surrogate modeling, and computational fluid dynamics (CFD) to optimize the aerodynamic characteristics of a prototype vehicle for Shell Eco Marathon, a global-level competition in which collegiate teams compete to make the most energy-efficient vehicles. It has 2 different vehicle classes - (1) Urban Concept and (2) Prototype. Urban concept cars are required to resemble the original on-road cars, whereas prototypes can have any design (must be compliant with Shell Eco Marathon official rules) whether 3 or 4-wheeled which is aimed at achieving ultimate performance in terms of mileage. By comparing it with conventional hit-and-trial design processes involving repeatedly modifying the vehicle body and carrying out simulations, the superiority of this method is demonstrated in achieving enhanced efficiency, reduced drag, and improved overall performance.

The present study breaks down the overall vehicle geometry into two segments- the side and the top profile. These profiles are composed of three-degree Bezier curves, the top profile being symmetric and the side profile having an asymmetric shape about the longitudinal axis. The control points of these curves are used as the parameters of the optimization study with the space for mounting the powertrain and steering components providing constraints on the value of these parameters. In the next step, a Response Surface Methodology is employed to circumvent the large number of simulations by creating a response surface between the control points and the objective function (Drag Coefficient). After this, the genetic algorithm is employed to ensure a globally optimal airfoil shape. This is achieved through probabilistic selection based on fitness, combining genetic information through crossover, introducing diversity through mutation, and iteratively evolving a population of potential solutions. The obtained results are analyzed and presented in the later stages of the research paper through aerodynamic visualizations and velocity/pressure distribution graphs of selected shapes of the vehicle.

## 2    Methodology

### 2.1    Numerical Model Validation

In the current work, wind tunnel data for Ahmed body [9] is utilized as a reference to validate the CFD model. The Ahmed body is a well-established benchmark used for validating CFD solvers, as it has all the necessary flow features encountered by modern automotive designs. Since the current study aims to propose a new vehicle design, there is no experimental data for validating the same model. Therefore, steady-state RANS simulations with the k-ω SST model [10] carried out using the commercial CFD code Ansys Fluent version 22.1 are utilized to first compute drag coefficients for the Ahmed Body, and in the next step, the validated model would be used for generating the required data for the optimization algorithm. The dimensions of the Ahmed body used for the study are presented in Figure 1. For the simulations, a freestream velocity of 60 m/s is assigned at the inlet corresponding to a Reynolds number of $4.25 \times 10^6$ based on the model length.

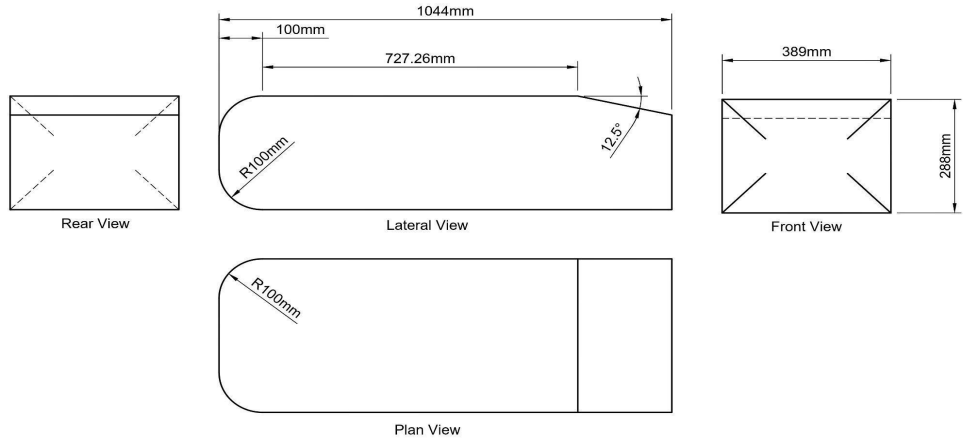

Figure 1: Ahmed Body dimensions

To examine the sensitivity of the numerical solution on the grid size, a grid independence test is conducted and the results for each grid are compared with the experimental data in Table 1 shown below.

Table 1: Results of Grid Independence and Validation study for the Ahmed Body

| Parameter | Coarse | Medium | Fine | Experiment |
|---|---|---|---|---|
| Cd [-] | 0.297 | 0.268 | 0.248 | 0.230 |
| Cell Count [-] | $0.9 \times 10^6$ | $1.4 \times 10^6$ | $1.8 \times 10^6$ | - |
| Error [%] | 29.13 | 16.52 | 7.82 | - |

The accuracy of the numerical model can be clearly observed from Table 1. The fine grid is able to capture the quantity of interest, i.e., drag coefficient ($C_d$) accurately, and hence would be used for further parts of the study.

## 2.2  Grid Convergence Study

To ensure accuracy as well as efficiency in the simulations, a high-quality Polyhex core grid, shown in Figure 2(a), is used for discretization of the flow domain. The origin of the domain is at the nose of the vehicle body and the inlet is kept at a distance of 8 times the vehicle length and the outlet at a distance of 20 times the vehicle length. The side and the top walls are kept at a distance of 8 times the vehicle length. To save computational resources, finer cells are used to model the wake of the vehicle, and relatively coarser cells are used in the surrounding regions, For capturing the effect of the boundary layer, 20 prismatic cells are used with a first layer height of 1mm and a growth rate of 1.12. The gradient discretization is done using the least square cell-based method and the second-order upwind scheme is used for spatial discretization of other transport quantities. The convergence criteria for residuals is set to $10^{-6}$ and a total of 1000 iterations are used to ensure the convergence of residuals and the drag coefficient.

To ensure solution independence with respect to the grid count, a grid convergence study is carried out using the Richardson extrapolation method [11]. The results of the study are presented in Figure 2(b). Three different grids were used with cell counts similar to Table 1. It can be observed that the solution for the fine grid is very close to the extrapolated solution for an infinitely fine grid. The error metrics for the grid are as follows: extrapolated relative error is 0.91%, relative error is 3.61%, and finally, the grid convergence index for the medium and the fine grid is 1.13.

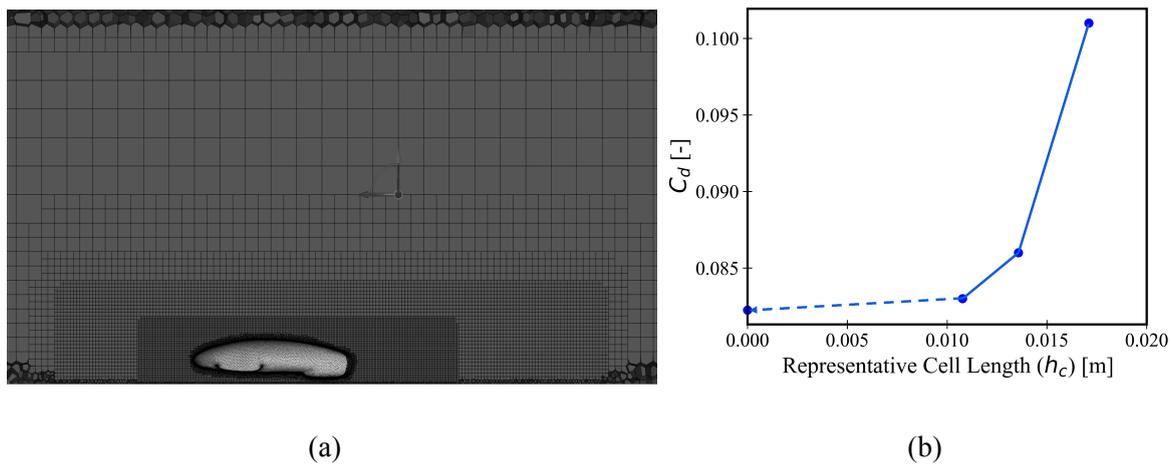

(a)          (b)

Figure 2: (a) Computational Grid, (b) Variation of $C_d$ and Representative Cell Length

## 2.3  Parametrization of Airfoils

Parameterization of airfoils refers to the process of representing the complex geometry of an airfoil using a set of numerical parameters that are adjusted to satisfy an objective from the desired shape. These parameters are used to define the shape of the airfoil mathematically, allowing engineers and researchers to describe, analyze, and manipulate the airfoil's geometry in a systematic and standardized manner. The process is essential in the field of aerodynamics and vehicle design because it simplifies the representation of airfoil shapes, making it easier to perform computational simulations, conduct optimizations, and compare different airfoil designs.

Parametrization of airfoils is crucial for several reasons in the field of the development of aerodynamically efficient design. It allows engineers to optimize the shape of airfoils for specific aerodynamic performance metrics, such as lift-to-drag ratio. In our case, the drag coefficient serves as the foundational performance metric. The process allows researchers to study the effect of varying the different parameters on the behavior and structure of the object that is being designed thereby helping the designers in their decision-making process. Parametric models and techniques foster innovation and research by allowing designers and researchers to experiment with unconventional airfoil shapes. The efficiency of the technique is also extremely crucial as efficient parameterization methods help in quickly exploring a wide range of airfoil shapes, enabling engineers to identify the most efficient designs for various applications. The effect of airfoil shape parameterization in airfoil optimum shape and its convergence rate has been investigated by [12] and it was shown that a suitable parameterization method can improve the convergence rate of the optimization algorithm.

Among the various methods available for the process of parametrization, smooth parametrization of the airfoils is accomplished using B-spline curves. The Bezier curve is found to be suitable for airfoil parameterization, and the shape of the upper and lower curves can be precisely and visually controlled by Bezier control points [13]. Based on the vehicle's design constraints, the airfoils' chord is maintained at 3.2 meters. As the airfoil's lower surface represents the vehicle's floor in the side profile, it is left unchanged. Three control points, as seen in Figure 3, govern the shape of the top and side profiles. Only one side of the airfoil contains control points since the top profile is kept symmetrical about the airfoil's chord; the other side is produced by mirroring the top portion about the chord line, as illustrated in Figure 3 (b). The x-coordinates of the control points are fixed for both profiles. Table 2 lists the normalized values of the upper and lower limits for each control point. These values were decided based on the space constraints of other vehicle components, like the steering assembly, powertrain, and the space occupied by the driver. These control points are varied through various iterations between the upper and the lower limits to search for an optimum design with minimum $C_d$.

Table 2: Limits of control points for side and top profiles of the vehicle

| Point ID/Constraints | X/c | Y/c Lower | Y/c Upper |
|---|---|---|---|
| **P1** | 0.1250 | 0.0625 | 0.1875 |
| **P2** | 0.5000 | 0.1094 | 0.1875 |
| **P3** | 0.8750 | 0.0625 | 0.1875 |
| **P4** | 0.3125 | 0.1093 | 0.1406 |
| **P5** | 0.5312 | 0.0937 | 0.1406 |
| **P6** | 0.6875 | 0.0625 | 0.1406 |

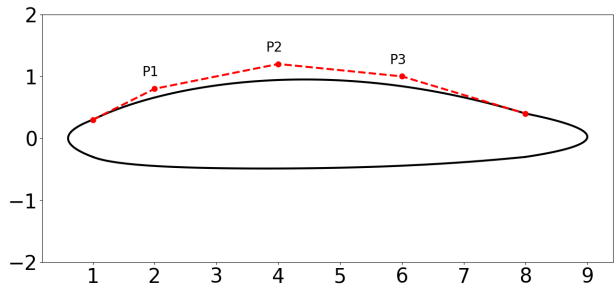

(a)

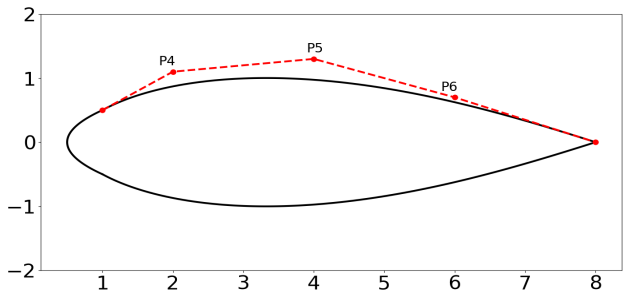

(b)

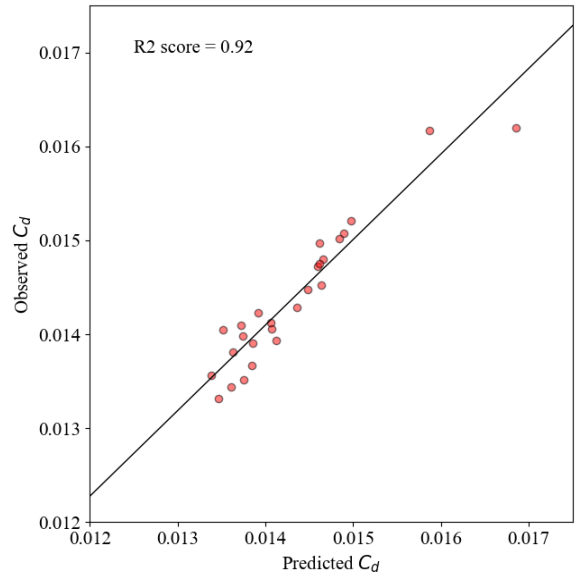

(c)

Figure 3: Control Points of (a) side profile (b) top profile (c) Validation of Kriging surrogate model with ground truth CFD data

## 2.3 Response Surface Methodology

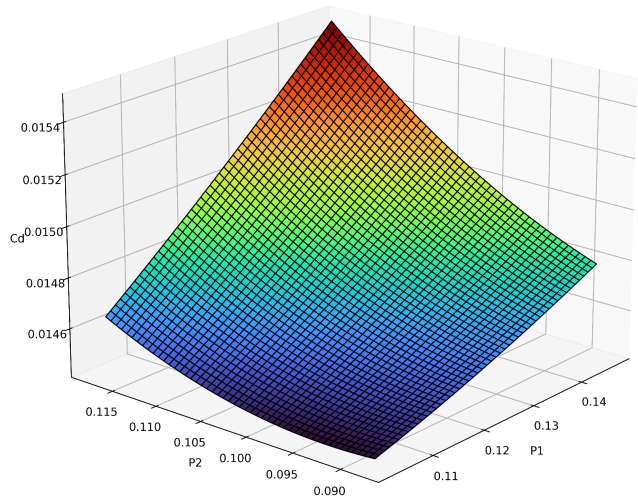

Figure 4: Response Surface Generated by Kriging Surrogate Model for Top Profile for P3 = 0.08

Response Surface Methodology (RSM) is used in problems where a large number of simulations are required for determining the optimal solution. In the present study, for the top and side views of the vehicle, the positions of two control points, each having an abscissa and an ordinate, will have to be optimized in order to obtain the most efficient design. Considering all the possible airfoils that can be generated, this leads to a large number of samples and it is impractical and inefficient to run thousands of CFD simulations. Thus, RSM is used to construct a response surface between the independent variables (control points) and the objective function (drag coefficient). Surrogate models are employed in RSM to construct the response surface. Kriging surrogate model is chosen for this objective. Kriging is an interpolation technique that involves forming a linear combination of a known function $f_i(\mathbf{x})$ and incorporating it with a realization of a stochastic process $Z(\mathbf{x})$

$$\sum_{i=1}^{k} \beta_i f_i(x) + Z(x)$$

where $Z(\mathbf{x})$ is a realization of a stochastic process with mean zero given by

$$cov\left[Z(x^{(i)}), Z(x^{(j)})\right] = \sigma^2 R(x^{(i)}, x^{(j)})$$

where $\sigma^2$ is the process variance, and R is the correlation.

The accuracy of the surrogate model is shown in Fig. 3 (c).

## 2.4   Optimization Algorithm - The Genetic Algorithm (GA)

Evolutionary algorithms (EAs) are a family of optimization algorithms inspired by the principles of biological evolution. They are part of the broader class of evolutionary computation techniques, which also include genetic algorithms, genetic programming, evolutionary strategies, and differential evolution, among others. EAs are used to find approximate solutions to complex optimization and search problems, especially in cases where the search space is large, multidimensional, or poorly understood. Evolutionary algorithms such as GA are used to obtain the globally optimum solution from the design space. The GA method imitates natural evolution wherein an initial population is chosen before crossover and mutation operations are often carried out depending on each individual's score for the objective function [14,15]. The genetic algorithm technique can be employed efficiently and accurately to produce globally optimal airfoils with excellent aerodynamic properties using desired objective values such as that of lift coefficient, drag coefficient, or the ratio of the two [16]. The primary objective of this study is the minimization of the net drag force experienced by the vehicle by optimizing its shape. The mathematical principles involve probabilistic selection based on fitness, combining genetic information through crossover, introducing diversity through mutation, and iteratively evolving a population of potential solutions. Over generations, the algorithm refines the solutions, converging towards an optimal or near-optimal solution for the given problem. The choice of appropriate parameters, such as selection methods, crossover points, and mutation rates, significantly influences the performance of the genetic algorithm. Intelligent and advanced GA also exist in the state-of-the-art that incorporates a fractional factorial crossover and SA mutation, which effectively improves the evolutionary efficiency of the conventional GA [17].

The process starts with "initialization" i.e. generating an initial population of candidate airfoil shapes by randomly generating or selecting individuals from the search space defined by the parameterization process.

$$x = (x_1, x_2, x_3, \ldots..)$$

Further, the performance of each individual is evaluated using appropriate CFD simulations or analytical methods. The fitness value for each profile is evaluated based on the defined objective (drag minimization of prototype vehicle). The fitness function f(x) evaluates the quality of a solution represented by a chromosome x. It assigns a numerical value indicating how well the solution solves the problem. Based on the fitness values, individuals from the population are selected to serve as parents for the next generation.

$$P(x) = \frac{f(x)}{\Sigma f(xi)}$$

Individuals with a high fitness probability (lower fitness score) will be more likely to be chosen. Crossover and mutation are two genetic operations that are used to produce offspring from the chosen parents. While mutation causes random changes to the offspring, crossover entails merging the genetic information from two parents. The parents and offspring from the previous steps are combined to create a new population with random new samples that could better fit the objective function. The optimization process terminates on reaching a maximum number of generations or achieving a satisfactory solution depending on the level of fitness at the optimum solution. The best individual in the final population represents the optimized airfoil shape. The comprehensive flow of operation of the genetic algorithm has been represented in the flowchart in Figure 5.

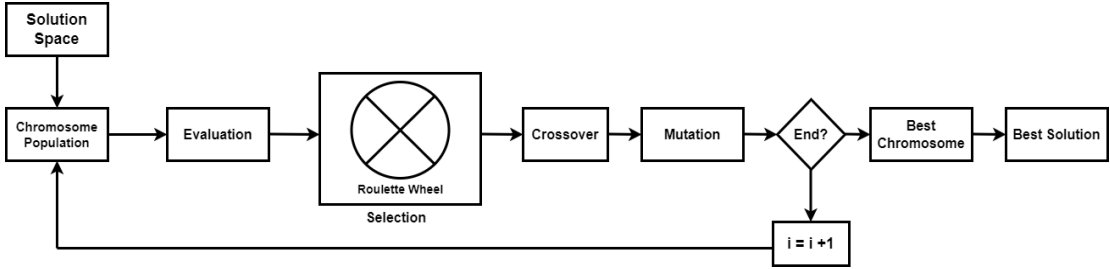

Figure 5: Genetic algorithm process flowchart

# 3 Results

## 3.1 Optimum Parameters

The results of the study are summarized in Table 3. Here initialized design refers to the vehicle body designed using the initial side and top profiles fed into the optimization algorithm and the optimum design refers to the final vehicle body designed using the optimized side and top profiles. An improvement of 26.6% can be observed in the overall Drag Coefficient and an improvement of 51.1 % in the Drag Area demonstrating the effectiveness of the Genetic algorithm in carrying out multi-parameter optimization studies. The optimum and initialized design are presented in Figure 6.

Table 3: Comparison of Initialized and Optimized Vehicle Design

| Design | Drag Coefficient ($C_d$) [-] | Frontal Area (A) [$m^2$] | Drag Area ($C_d \cdot A$) [$m^2$] |
|---|---|---|---|
| **Initialised Design** | 0.094 | 0.72 | 0.0677 |
| **Optimum Design** | 0.069 | 0.48 | 0.0331 |

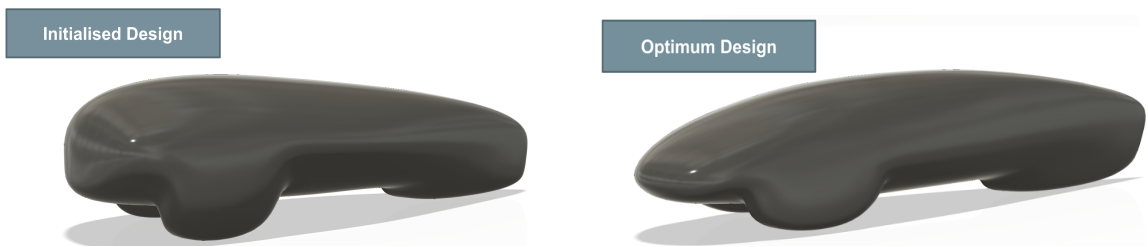

Figure 6: Comparison between the Initial and Optimized design

The graph in Figure 7 (a) provides insights into how the optimization process is progressing over time and is used to compare the convergence histories throughout the optimization process. The objective function represents the quality or fitness of the solutions generated by the algorithm and the number of iterations refers to the generations the algorithm has gone through. As the number of iterations increases, the objective function achieves a steady value. The rate of convergence towards the optimum solution can also be inferred from the graph showing the efficiency of the algorithm to obtain optimum solutions. At the end of the optimization process, the objective function value stabilizes and this final value gives an idea of the quality of the solution that the genetic algorithm has found. In addition, the graph in Figure 7 (b) represents the convergence monitor for $C_d$ for both the initial and optimum design as given by the CFD solver.

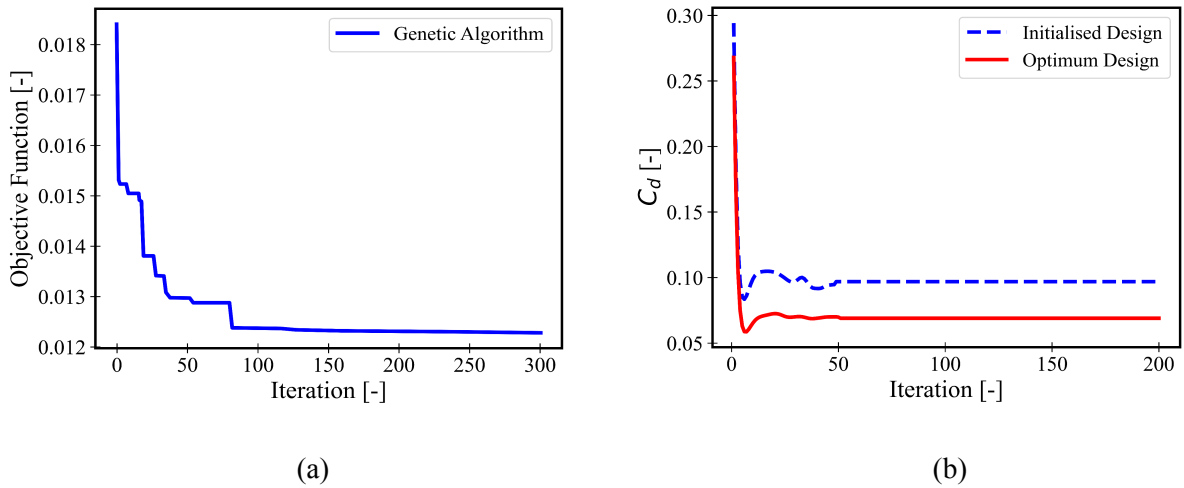

(a)  (b)

Figure 7: (a) Convergence of Objective Function for top profile using GA, (b) Convergence of $C_d$ monitor obtained from the RANS simulation of initialized and optimum Design

## 3.2 Velocity Distribution

The velocity distribution around the initial and optimum geometry are presented side by side for comparison in Figure 8. The incoming air from the left-hand side impinges upon the nose of the vehicle and stagnates. The stagnation region is more prominent for the initial geometry due to its bluff shape and steep change in the curvature. On the other hand, the incoming air striking the nose of the optimum design flows smoothly over it due to its gradually changing curvature leading to a relatively smaller stagnation region as evident in Figure 8. Also, the wake of the initial geometry is more pronounced, implying a greater momentum deficit and thus larger drag force pulling the vehicle to the leeward side. The larger momentum deficit in the wake and a more prominent stagnation zone increase the imbalance of pressure in the longitudinal direction leading to a larger form drag for the initialized design which is also evident from Table 3.

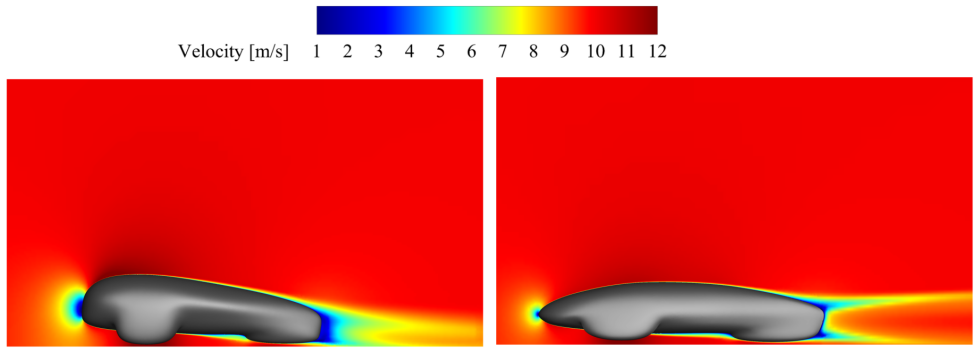

(a)

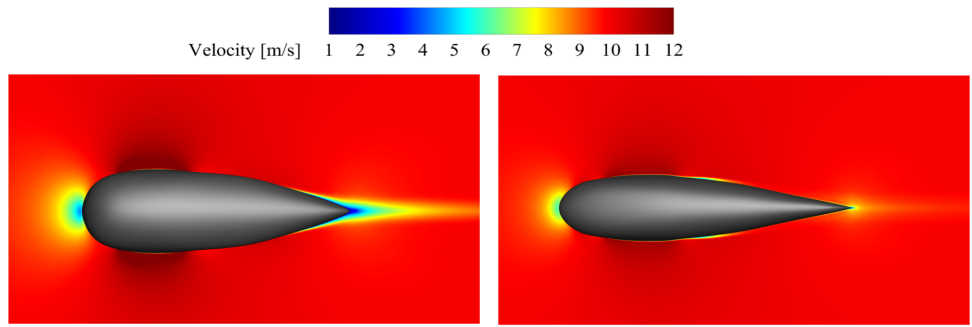

(b)

Figure 8: Velocity Distribution around the vehicle (a) Side View, (b) Top View

## 3.2  Pressure Distribution

A comparison has been made in the pressure distribution around both the geometries in Figure 9. The larger stagnation zone for the initial geometry as discussed previously, leads to a larger static pressure at the nose of the vehicle. This buildup of incoming air is mitigated for the optimum design due to its streamlined nose which aids the incoming air to flow over it easily without much hindrance. As the incoming air strikes the nose of the vehicle and starts to flow it, the pressure decreases due to the increase in the flow velocity due to the convexity of the curvature. As the air reaches the top of the vehicle and starts to flow down the curvature towards the rear end, the pressure starts to increase. This creates a pressure differential along the longitudinal direction contributing towards the form drag or pressure-induced drag force. Due to the symmetry of curvature for the optimum design as evident from the side view shown in Figure 9 (a), there is a greater pressure recovery at the rear end of the vehicle resulting in a much lower $C_d$ compared to the initialized geometry.

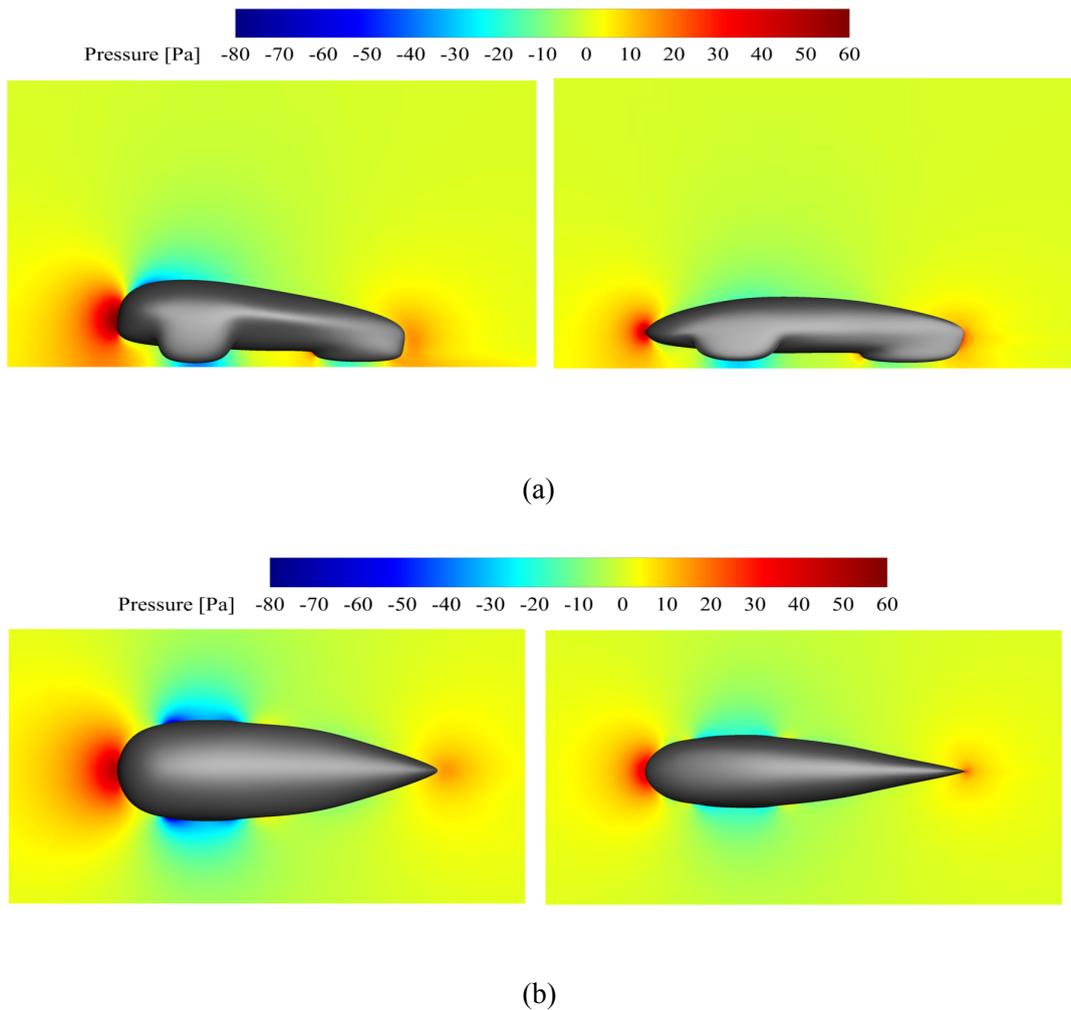

Figure 9: Pressure Distribution around the vehicle (a) Side View, (b) Top View

## 4  Conclusion and Future Work

In conclusion, the proposed study demonstrates the effectiveness of utilizing Genetic Algorithm to optimize the design of a fuel-efficient prototype vehicle. This is achieved by employing parametrization to simplify simulations and a response surface methodology to create a response surface. Further, genetic algorithms are used to obtain globally optimum solutions. The minimization of drag coefficient through this approach led to a significant improvement in aerodynamic efficiency. These findings highlight the potential of the present methodology in achieving enhanced performance and fuel economy through improvements in vehicle design. The current approach only focuses on the side and top profile, but to cater to design optimization for more complex geometries further developments can be done in the area of parameterization of three-dimensional geometries.